\documentclass{article}

\date{\today}

\newcommand{\insertplot}[5]{\begin{figure}
 \hfill\hbox to 0.05in{\vbox to #5in{\vfill
 \inputplot{#1}{#4}{#5}}\hfill}
 \hfill\vspace{-.1in}
 \caption{#2}\label{#3}
 \end{figure}}
 \newcommand{\inputplot}[3]{
 \special{ps: plotfile #1}
}
\newcounter{fig}   

\usepackage{epsfig}
\usepackage{amsmath}
\usepackage{amsfonts} 
\usepackage{graphicx}
\usepackage[german, english]{babel}
\usepackage{a4wide}
\usepackage{amsmath}
\usepackage{amssymb}
\usepackage{ifthen}
\usepackage{epsfig}

\begin{document}

\newcommand{\dd}{\mbox{d}}
\newcommand{\tr}{\mbox{tr}}
\newcommand{\la}{\lambda}
\newcommand{\ta}{\theta}
\newcommand{\f}{\phi}
\newcommand{\vf}{\varphi}
\newcommand{\ka}{\kappa}
\newcommand{\al}{\alpha}
\newcommand{\ga}{\gamma}
\newcommand{\de}{\delta}
\newcommand{\si}{\sigma}
\newcommand{\bomega}{\mbox{\boldmath $\omega$}}
\newcommand{\bsi}{\mbox{\boldmath $\sigma$}}
\newcommand{\bchi}{\mbox{\boldmath $\chi$}}
\newcommand{\bal}{\mbox{\boldmath $\alpha$}}
\newcommand{\bpsi}{\mbox{\boldmath $\psi$}}
\newcommand{\brho}{\mbox{\boldmath $\varrho$}}
\newcommand{\beps}{\mbox{\boldmath $\varepsilon$}}
\newcommand{\bxi}{\mbox{\boldmath $\xi$}}
\newcommand{\bbeta}{\mbox{\boldmath $\beta$}}
\newcommand{\ee}{\end{equation}}
\newcommand{\eea}{\end{eqnarray}}
\newcommand{\be}{\begin{equation}}
\newcommand{\bea}{\begin{eqnarray}}
\newcommand{\ii}{\mbox{i}}
\newcommand{\e}{\mbox{e}}
\newcommand{\pa}{\partial}
\newcommand{\Om}{\Omega}
\newcommand{\vep}{\varepsilon}
\newcommand{\bfph}{{\bf \phi}}
\newcommand{\lm}{\lambda}
\def\theequation{\arabic{equation}} 
\renewcommand{\thefootnote}{\fnsymbol{footnote}}
\newcommand{\re}[1]{(\ref{#1})}
\newcommand{\R}{{\rm I \hspace{-0.52ex} R}}
\newcommand{\N}{{\sf N\hspace*{-1.0ex}\rule{0.15ex}%
{1.3ex}\hspace*{1.0ex}}}
\newcommand{\Q}{{\sf Q\hspace*{-1.1ex}\rule{0.15ex}%
{1.5ex}\hspace*{1.1ex}}}
\newcommand{\C}{{\sf C\hspace*{-0.9ex}\rule{0.15ex}%
{1.3ex}\hspace*{0.9ex}}}
\newcommand{\eins}{1\hspace{-0.56ex}{\rm I}}
\renewcommand{\thefootnote}{\arabic{footnote}}

\title{ $d=4+1$ gravitating 
nonabelian solutions\\
with bi-azimuthal  symmetry }

\author{{\large Eugen Radu}$^{\dagger\star}$, {\large Ya. Shnir}$^{\ddagger}$
 and {\large D. H. Tchrakian}$^{\dagger\star}$ \\ \\
$^{\dagger}${\small Department of Mathematical Physics, National
University of Ireland Maynooth,} 
\\ {\small Maynooth, Ireland}
\\ $^{\star}${\small School of Theoretical Physics -- DIAS, 10 
Burlington Road, Dublin 4, Ireland } 
\\ $^{\ddagger}${\small Institut f\"ur Physik, Universit\"at Oldenburg,
Postfach 2503 D-26111 Oldenburg, Germany}
}

\maketitle

\begin{abstract}
We construct static, asymptotically flat solutions of  $SU(2)$ Einstein-Yang-Mills theory
in $4+1$ dimensions, subject to bi-azimuthal symmetry. 
Both particle-like and black hole solutions are considered for two different
sets of boundary conditions in the Yang--Mills sector, corresponding
to multisolitons and soliton-antisoliton pairs. For gravitating multi-soliton solutions, 
we find that their mass per unit charge is lower than the mass of the
corresponding unit charge, spherically symmetric soliton.
\end{abstract}

%%%%%%%%%%%%%%%%%%%%%%%%%%%%%%%%%%%%%%%%%%%%%%%%%%%%%%%%%%%%%%%%%%
\section{Introduction}
%%%%%%%%%%%%%%%%%%%%%%%%%%%%%%%%%%%%%%%%%%%%%%%%%%%%%%%%%%%%%%%%%%

The last years have seen an increasing interest in the solutions of 
Einstein equations involving more than four dimensions.
The  results in the literature indicate that the
physics in higher-dimensional general relativity
is far richer and complex than in the standard four-dimensional theory.

Naturally, most of the studies in the literature were carried out for 
vacuum solutions or to configurations with an Abelian matter content.
At the same time, a number of results in the literature clearly
indicate that solutions to the Einstein equations coupled to non Abelian
matter fields possess a much richer structure than in the U(1) 
case (see \cite{Volkov:1998cc} for a survey 
of the situation in four dimensions and the more recent 
review~\cite{Volkov:2006xt} for $d>4$), most
notably in that they are not restricted to black holes, but can also be regular.

Physically reasonable stationary 
vacuum solutions in higher dimensional spacetimes, $d\ge 4$, 
 fall in two categories, distinguished by their asymptotic behaviours.
In the first category, there are the static spherically symmetric solutions generalising the
 $d=4$ Schwarzschild black hole, found by Tangherlini a long time ago~\cite{Tangherlini:1963bw}, 
the rotating Myers-Perry solution  \cite{Myers:1986un}
generalising the four dimensional Kerr black hole, and more recently the black ring solutions 
\cite{Emparan:2001wn, Elvang:2007rd}.
In all these cases, the $d$-dimensional spacetime approaches 
asymptotically the ${\cal M}^{d}$ Minkowski background.
The second category are the black string solutions,
and the corresponding black $p-$branes generalizations~\cite{Horowitz:1991cd}.
The  black strings  approach asymptotically
$d-1$ dimensional Minkowski-spacetime times a circle, ${\cal M}^{d-1}\times S^1$,
and in the simplest case present translational symmetry along the 
extra-coordinate direction. (Such configurations are important if one supposes 
the existence of extra dimensions in the universe, 
which are likely to be compact and described by a Kaluza-Klein (KK) theory.)

As is the case with the usual Schwarzschild black hole, 
all these vacuum solutions  can be extended to
describe  configurations with an Abelian matter content. 
The inclusion of non Abelian
matter fields is less systematic and is complicated by the fact 
that all known such solutions can only be
evaluated numerically, starting from the earliest found 
Einstein--Yang-Mills (EYM) solution in four
spacetime dimensions discovered by Bartnik and McKinnon~\cite{Bartnik:1988am}.

Spherically symmetric solutions to EYM systems in $d$-spacetime dimensions, 
approaching asymptotically the ${\cal M}^{d}$ Minkowski background, were
constructed systematically in
\cite{Brihaye:2002hr}-\cite{Brihaye:2006xc}.
The Yang--Mills (YM) sector of the systems studied there consisted of 
all needed terms belonging to the YM
hierarchy~\cite{Tchrakian:1984gq,hier}, which 
are higher order in the YM curvature in the manner of the
Skyrme model. 
Such terms may arise in the low energy    effective  action of string
theory~\cite{Tseytlin,BRS,Cederwall:2001bt}. 
It has been established that only in the presence of these
higher order in the YM curvature terms, does the EYM solution lead to a finite mass. 
In the absence of
such Skyrme-like terms, 
for example in \cite{Volkov:2001tb, Okuyama:2002mh} (in $d=5$), the mass of the solution diverges.
Both particle like and black hole solutions were constructed.
The properties of these configurations are rather
different from the familiar Bartnik-McKinnon 
solutions~\cite{Bartnik:1988am} in $d=4$, and are
somewhat more akin to the gravitating monopole 
solutions to EYM-Higgs system \cite{Breitenlohner:1992}, which
is not surprising since the latter features 
the dimensionful vaccuum expectation value (VEV) of the Higgs field, while the former
contain additional dimensionful terms entering as the couplings of the higher order YM terms.

As for solutions to the EYM system in $d$ dimensional 
spacetime whose vacuum has the structure of
 ${\cal M}^{d-1}\times S^1$ like the black string solutions, 
these are only constructed
if one of the spacelike dimensions is supposed to be compact, 
and a Kaluza-Klein descent is performed, essentially eliminating
that coordinate. Such solutions are given  for $d=5$ in
\cite{Volkov:2001tb}-\cite{Brihaye:2006ws}.
However, in the present work we
will not be concerned with this type of solutions.

Our aim in the present work is to extend the construction 
of asymptotically flat finite mass EYM solutions
vacuum, relaxing the constraint of spherical symmetry 
in the $d-1$ dimensional spacelike subspace as in
\cite{Brihaye:2002hr}-\cite{Brihaye:2006xc}.

The simplest possibility is to consider the 
imposition of a symmetry which leads to two a dimensional
reduced effective system, rather than the one dimensional one in the previous examples.
This is the first such attempt in the literature, and 
the numerical work of solving a two dimensional
EYM boundary value problem is a task of considerable complexity. 
To achieve a two dimensional subsystem, we have
found that the simplest option is to impose bi-azimuthal symmetry 
on the $d=5$ static EYM system. 
This is why we
have restricted to $d=5$, for otherwise a similar 
application of azimuthal symmetries in each plane would
result in multi-azimuthal~\footnote{If one applied instead, 
spherical symmetry in the $d-2$ dimensional
subspace of the $d-1$ spacelike dimensions, then the 
residual subsystems will always be {\it two dimensional}
irrespective of the value of $d$. For example in $d=5$ this 
would be the $SO(3)$ symmetry in the $3$ dimensional
subspace of the $3$ dimensional subspace of the $4$ spacelike 
dimensions, exactly as for the axially symmetric
instantons~\cite{Witten:1976ck}. While this may appear to be 
an attractive alternative, we have found that
tackling the boundary value problem in that case is a 
considerably harder task, even in $d=4$.} subsystems,
with higher dimensional boundary value problems to be solved, 
technically beyond the scope of this work.
Indeed, as a warmup for the task at hand, we have carried 
out the same program in \cite{Radu:2006qf} recently,
with the dilaton replacing gravity.

While we have restricted to five dimensional EYM solutions for 
technical reasons, this example is of considerable
physical relevance since it enters all 
$d=5$ gauged supergravities as the basic
building block and one can expect the basic 
features of its solutions to be generic. 
Also special about $d=5$
gravitating YM is the particular critical properties of the solutions 
present in all $d=4p+1$ analysed in
\cite{Breitenlohner:2005hx}, and first discovered in \cite{Brihaye:2002jg}. 
Indeed in the $d=5$ YM-dilaton (YMd)
system, studied in \cite{Radu:2006qf}, these critical 
properties were present, providing yet another confirmation
that dilaton interactions with YM, mimic~\cite{Maison:2004hb} those with Einstein gravity.
 
The purpose of this paper is to present numerical arguments for the existence of a class of static 
$d=5$ solutions to the EYM equations of the model studied in \cite{Brihaye:2002jg}, but now,
subject to bi-azimuthal symmetry. 
These configurations present a spacetime
symmetry group $R\times U(1)\times U(1)$, 
where $R$ denotes time translation symmetry and the $U(1)$ factors
the rotation symmetry in two orthogonal planes. 
We present both regular and black hole solutions. In
the particle like case we find solutions with many similar properties to those 
of the four dimensional SU(2) YM multi-instantons and
composite instanton-antiinstanton bound states with $ U(1)\times U(1)$ symmetry, reported in
Ref.~\cite{Radu:2006gg}. Dilatonic generalizations of these solutions have been considered in 
\cite{Radu:2006qf}, in which higher order gauge curvature terms were included
in the action to enable the existence of finite mass solutions.

%%%%%%%%%%%%%%%%%%%%%%%%%%%%%%%%%%%%%%%%%%%%%%%%%%%%%%%%%%%%%%%%%%
\section{The model}
%%%%%%%%%%%%%%%%%%%%%%%%%%%%%%%%%%%%%%%%%%%%%%%%%%%%%%%%%%%%%%%%%%
%%%%%%%%%%%%%%%%%%%%%%%%%%%%%%%%%%%%%%%%%%%%%%%%%%%%%%%%%%%%%%%%%%
\subsection{The ansatz and field equations}
%%%%%%%%%%%%%%%%%%%%%%%%%%%%%%%%%%%%%%%%%%%%%%%%%%%%%%%%%%%%%%%%%%
We consider the five dimensional SU(2)  EYM  action
\begin{eqnarray}
\label{grav}
S=\int d^5x\sqrt{-g}~\left(\frac{R}{16 \pi G}-{\cal L}_m\right),
\end{eqnarray}
where ${\cal L}_m$ is given by the superposition of the
$p=1$ and $p=2$ terms in the YM hierarchy \cite{Brihaye:2002jg}
\be
\label{L}
{\cal L}_m= 
 \frac{\tau_1}{2\cdot 2!} \mbox{Tr}\,
{\cal F}_{\mu \nu}^2+
\frac{\tau_2}{2\cdot 4!}  \mbox{Tr}\,{\cal F}_{\mu \nu \rho \sigma}^2~, 
\ee
with ${\cal F}_{\mu\nu}=\pa_{\mu}{\cal A}_{\nu}-\pa_{\mu}{\cal A}_{\nu}+
i[{\cal A}_{\mu},{\cal A}_{\nu}]$ the $2$-form YM curvature and
${\cal F}_{\mu \nu \rho \sigma}=\{{\cal F}_{\mu[\nu},{\cal F}_{\rho \sigma]}\}$
 the $4$-form YM curvature consisting of the totally antisymmetrised
product of two YM $2$-form YM curvatures  (the bracket $[\nu\rho\si]$
implies cyclic symmetry). $\tau_1$ and $\tau_2$ are dimensionfull coupling
strengths.

Variation of the action (\ref{grav})
 with respect to the metric $g^{\mu\nu}$  and gauge potential $A_\mu$
leads to the EYM equations
\begin{eqnarray}
\label{einstein-eqs}
R_{\mu\nu}-\frac{1}{2}g_{\mu\nu}R    = 8 \pi G~(T_{\mu\nu}^{(1)}+T_{\mu\nu}^{(2)}),
\\
\label{YM-eqs}
\tau_1D_{\mu}{\cal F}^{\mu\nu} +
\frac12\tau_2\{{\cal F}_{\rho\si},D_{\mu} 
{\cal F}^{\mu\nu\rho\si} \}=0,
\end{eqnarray}
where  
\be
T_{\mu\nu}^{(p)}=
\mbox{Tr}\{ {\cal F}(2p)_{\mu\lambda_1\lambda_2...\lambda_{2p-1}}
{\cal F}(2p)_{\nu}{}^{\lambda_1\lambda_2...\lambda_{2p-1}}
-\frac{1}{4p}g_{\mu\nu}\ {\cal F}(2p)_{\lambda_1\lambda_2...\lambda_{2p}}
{\cal F}(2p)^{\lambda_1\lambda_2...\lambda_{2p}}\} ,
\label{pstress}
\ee
is the energy-momentum tensor for the $p-$th YM term in (\ref{L}), $p=1,2$.

We consider a $d=5$ static metric form with two orthogonal commuting
rotational Killing vectors
\begin{eqnarray}
\label{metric}
ds^2=-f(r,\theta)dt^2+\frac{s(r,\theta)}{f(r,\theta)}(dr^2+r^2d \theta^2)
+\frac{l(r,\theta)}{f(r,\theta)} r^2\sin^2 \theta d \varphi^2+
\frac{p(r,\theta)}{f(r,\theta)} r^2\cos^2 \theta d \psi^2~,
\end{eqnarray}
where $r$ is the radial coordinate, 
and $\theta,\varphi,\psi$ are Hopf coordinates in $S^3$, with
$0\leq \theta\leq \pi/2$
and $0\leq \varphi,\psi\leq 2 \pi$.

The construction of a YM Ansatz compatible with the symmetries of the above line element has been
discussed at length in \cite{Radu:2006qf}, \cite{Radu:2006gg}. 
The purely magnetic gauge connection has
six nonvanishing components and reads
\begin{eqnarray}
\label{RR-Aa}
&& {\cal A}=\frac{1}{2}u_3a_r(r,\theta)dr+\frac{1}{2}u_3a_\theta(r,\theta)d\theta
\\
\nonumber
&&{~~~~~}+\left(\frac{1}{2}u_1\chi^1(r,\theta)+\frac{1}{2}u_2\chi^2(r,\theta)+\frac{n}{2}u_3\right)d\varphi
+\left(\frac{1}{2}u_1\xi^1(r,\theta)+\frac{1}{2}u_2\xi^2(r,\theta)+\frac{n}{2}u_3\right)d\psi, 
\end{eqnarray}
where
$u_1=\sin n(\varphi+\psi)\sigma_1-\cos n(\varphi+\psi)\sigma_2,~
u_2=\cos n(\varphi+\psi)\sigma_1+\sin n(\varphi+\psi)\sigma_2,~
u_3=\sigma_3,$
$\sigma_i$ being the Pauli matrices 
and $n$  the winding number of the solutions, $n=1,2,\dots.$
In the flat space limit, the reduced action density
 describes a $U(1)$ Higgs like model 
with two effective Higgs fields $\chi^A$ and $\xi^A$ $(A=1,2)$, coupled minimally
to the $U(1)$ gauge connection $(a_{r},a_{\theta})$
\cite{Radu:2006gg}. 

To remove the
$U(1)$ residual gauge freedom of the connection,
we impose the usual gauge condition
%\be
%\label{gc}
$\partial_r a_r+\frac{1}{r}\partial_\theta a_\theta=0~.$
%\ee
%%%%%%%%%%%%%%%%%%%%%%%%%%%%%%%%%%%%%%%%%%%%%%%%%%%%%%%%%%%%%%%%%%
\subsection{Boundary conditions}
%%%%%%%%%%%%%%%%%%%%%%%%%%%%%%%%%%%%%%%%%%%%%%%%%%%%%%%%%%%%%%%%%%
In this paper we shall consider both globally regular and 
black hole solutions of the field equations 
(\ref{einstein-eqs}), (\ref{YM-eqs}).
The boundary conditions satisfied at infinity and 
at $\theta=0,\pi/2$ is the same in both cases, 
and are found from the requirements of finite 
energy and regularity of solutions.
At $r \to \infty $ one imposes  
\begin{eqnarray}
\label{rinfty}
a_r=0,~a_{\ta}=-2m,
~\chi^A=
(-1)^{m+1}n\left(
\begin{array}{c}
\sin 2m\ta \\
\cos 2m\ta
\end{array}
\right),~
\xi^A=-n\left(
\begin{array}{c}
\sin 2m\ta \\
\cos 2m\ta
\end{array}
\right)\,, 
~f=l=p=s=1~, 
\end{eqnarray}
with $m$ a  positive integer.
The following
boundary conditions holds for gauge potentials at $\theta=0$ 
\bea
a_r=\frac{1}{n}
\pa_{r}\xi^1,~~
a_{\ta}=\frac{1}{n}
\pa_{\ta}\xi^1,~~
\chi^1=0,~~
\xi^1=0,~~
\pa_{\ta}\chi^2=0,~~
\xi^2=-n~, 
\label{th0}
\eea
while for $\theta=\pi/2$ one imposes
\bea
\label{thp2}
a_r=\frac{1}{n}
\pa_{r}\chi^1,~~
a_{\ta}=\frac{1}{n}
\pa_{\ta}\chi^1,~~
\chi^1=0,~~
\xi^1=0,~~
\chi^2=-n,~~
\pa_{\ta}\xi^2=0.
\eea
The boundary conditions for the metric functions
at $\theta=0$ are 
$\partial_\theta f= \partial_\theta s=\partial_\theta l=\partial_\theta p=0, $
and agree with the boundary conditions on the $\theta=\pi/2$ axis.
There are also elementary flatness requirements which imposes
for the metric functions $s=l$ at $\theta=0$ and 
$s=p$ at $\theta=\pi/2$. 

To obtain globally regular EYM solutions with finite energy density we impose at 
the origin 
($r=0$) the boundary conditions
\be
\label{r01}
a_r=0~,~~ a_{\ta}=0~,~~
\chi^A=\left(
\begin{array}{c}
\ 0 \\
-n
\end{array}
\right)~,~~
\xi^A=\left(
\begin{array}{c}
\ 0 \\
-n
\end{array}
\right)\,~,~~
 \partial_r f= \partial_r s= \partial_r l= \partial_r p=0.
\ee
The black hole configurations possess an event horizon located at 
some constant value of the radial coordinate $r_h>0$,
where the following boundary conditions are imposed
\be
\label{r0}
a_r=0~,~\partial_r a_{\ta}=0~,~~
\partial_r \chi^A= 0~,~~
\partial_r \xi^A=0~,~~
 f=  s=  l=  p=0.
\ee
%The $m=1$  solutions present 
%a $n=1$ spherically limit with well-known properties
%\cite{Volkov:2001tb},\cite{Okuyama:2002mh}, \cite{Brihaye:2002jg}.
For $m=n=1$, these are the spherically symmetric solutions discussed in
 \cite{Volkov:2001tb},\cite{Brihaye:2002jg},\cite{Okuyama:2002mh}.
In this case the metric functions present no angular dependence, with 
$l=p=s$, while
$a_{\ta}=w(r)-1$, $
a_r=0,~\chi^1=-\xi^1=\frac{1}{2}(w(r)-1) \sin 2 
\theta,~ 
\chi^2=-(w(r)-1) \cos^2 \theta-1,~~\xi^2=-(w(r)-1) \sin^2 \theta-1.
$
%%%%%%%%%%%%%%%%%%%%%%%%%%%%%%%%%%%%%%%%%%%%%%%%%%%%%%%%%%%%%%%%%%
\subsection{Physical quantities}
%%%%%%%%%%%%%%%%%%%%%%%%%%%%%%%%%%%%%%%%%%%%%%%%%%%%%%%%%%%%%%%%%%

The mass $M$ of solutions is the conserved charge associated with
the Killing vector $v=\partial/\partial t$ and can be read from the asymptotic expression of the
$g_{tt}$-component of the metric tensor
\begin{eqnarray}
\label{f-exp}
-g_{tt}=f=1-\frac{8G M}{3\pi r^2}+O\big(\frac{1}{r^4}\big).
\end{eqnarray} 
The mass can also be expressed as an integral~\cite{Komar:1958wp} over 
the 3-sphere at spacelike infinity,
\begin{eqnarray}
\label{m1}
M= \frac{1}{16\pi G}\frac{3}{2}\oint_{\infty} v^{\mu;\nu}d^3 \Sigma_{\mu \nu}.
\end{eqnarray} 

The topological charge of the particle-like solutions as evaluated in \cite{Radu:2006gg} is
\be
\label{q}
q=\frac12\,[1-(-1)^m]n^2~,
\ee 
such that the  Pontryagin charge is nonzero only for odd $m$, being equal to $n^2$.
For even values of $m$, the solutions will describe soliton-antisoliton bound states.

To evaluate the Hawking temperature and entropy of the black hole solutions, we 
use the following expansions of the metric functions at the horizon
\begin{eqnarray}
\label{expan-h}
\nonumber
f(r,\theta)&=&f_2(\theta)\left(\frac{r-r_h}{r_h}\right)^2 
 + O\left(\frac{r-r_h}{r_h}\right)^3~,~~
p(r,\theta)=p_2(\theta)\left(\frac{r-r_h}{r_h}\right)^2 
 + O\left(\frac{r-r_h}{r_h}\right)^3\,
\\
\nonumber
l(r,\theta)&=&l_2(\theta)\left(\frac{r-r_h}{r_h}\right)^2 
 + O\left(\frac{r-r_h}{r_h}\right)^3\,
~~~
s(r,\theta)=s_2(\theta)\left(\frac{r-r_h}{r_h}\right)^2 
 + O\left(\frac{r-r_h}{r_h}\right)^3\,~.
\end{eqnarray} 
The zeroth law of black hole physics states that 
the surface gravity $\kappa$ 
is constant at the horizon of the black hole solutions,
where 
%\begin{equation}
%\label{kappa} 
$\kappa^2 =
-(1/4)g^{tt}g^{ij}(\partial_i g_{tt})(\partial_j g_{tt})\Big|_{r=r_h}. $
%\end{equation}
Since from general arguments the Hawking temperature $T_H$ is proportional
to the surface gravity
$\kappa $, $
T_H=\kappa /(2 \pi),
$
we obtain the relation
\begin{equation}
\label{temp}
T_H=\frac{f_2(\theta)}{2 \pi r_h \sqrt{s_2(\theta)} } 
\ . 
\end{equation}
One can show, with help of the $(r~\theta)$-component of the 
Einstein equations which implies
%\begin{eqnarray}
%\label{con}
$f_2  s_{2, \theta}=2s_2 f_{2,\theta},$
%\end{eqnarray}
that the temperature $T_H$, as given in (\ref{temp}), is indeed constant.

For the line element (\ref{metric}),
the  area $A$ 
of the event horizon  is given by 
\begin{equation}
\label{area}
A = 4\pi^2 r_h^3\int_0^{\pi/2}  d\theta \sin \theta \cos \theta
\sqrt{\frac{l_2(\theta)p_2(\theta) s_2(\theta)} {f_2^3(\theta)}}~.
\end{equation}
According to the usual thermodynamic arguments, the entropy $S$ is proportional 
to the area $A$, $ S = A/{4G}$.

We mention here also the Smarr-type relation
which follows from (\ref{m1}) together with Einstein equations
\begin{eqnarray}
\label{con}
\frac{2}{3}M=T_HS
-\frac{4\pi^2}{6}\int_{r_h}^{\infty}dr\int_{0}^{\pi/2}d\theta
\sin \theta \cos \theta
\frac{\sqrt{l p s}}{f} (T_t^t-\frac{1}{3}T).
\end{eqnarray}
This relation has been used in practice the
verify the accuracy of the numerical computation.

%%%%%%%%%%%%%%%%%%%%%%%%%%%%%%%%%%%%%%%%%%%%%%%%%%%%%%%%%%%%%%%%%%
\section{Properties of the solutions}
%%%%%%%%%%%%%%%%%%%%%%%%%%%%%%%%%%%%%%%%%%%%%%%%%%%%%%%%%%%%%%%%%%
The numerical calculations in this paper were performed by using the software 
package CADSOL, based on the Newton-Raphson method \cite{FIDISOL}. 
In this approach, the field equations are first discretised on a nonequidistant 
grid and the resulting system
is solved iteratively until convergence is achieved. In this scheme, a new radial variable $x=r/(1+r)$ 
(or $x=1-r_h/r$ for black hole solutions) is introduced which maps the semi-infinite region $[0,\infty)$ 
(or $[r_h,\infty)$) to the closed region $[0,1]$.
%%%%%%%%%%%%%%%%%%%%%%%%%%%%%%%%%%%%
\begin{figure}[h!]
\parbox{\textwidth}
{\centerline{
\mbox{
\epsfysize=15.0cm
\includegraphics[width=82mm,angle=0,keepaspectratio]{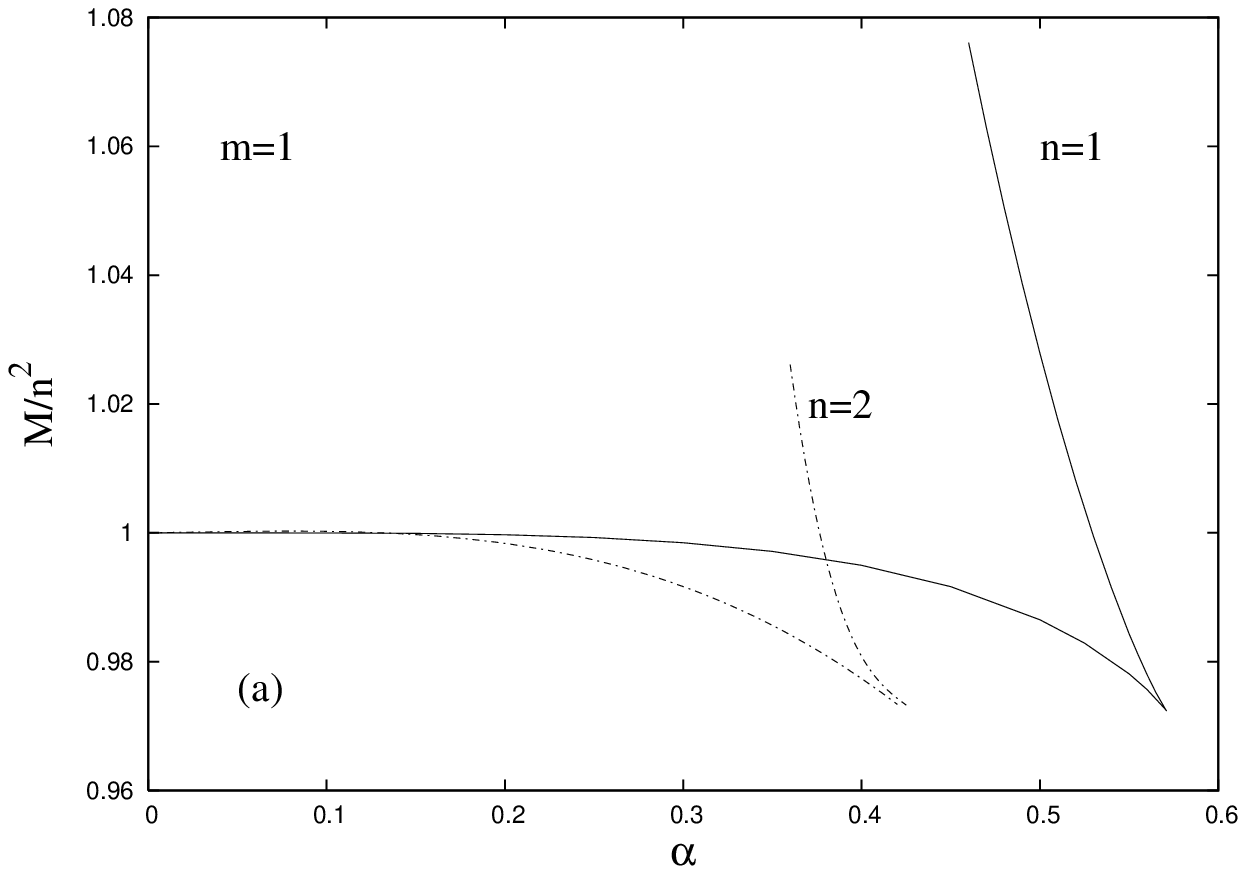}
\includegraphics[width=82mm,angle=0,keepaspectratio]{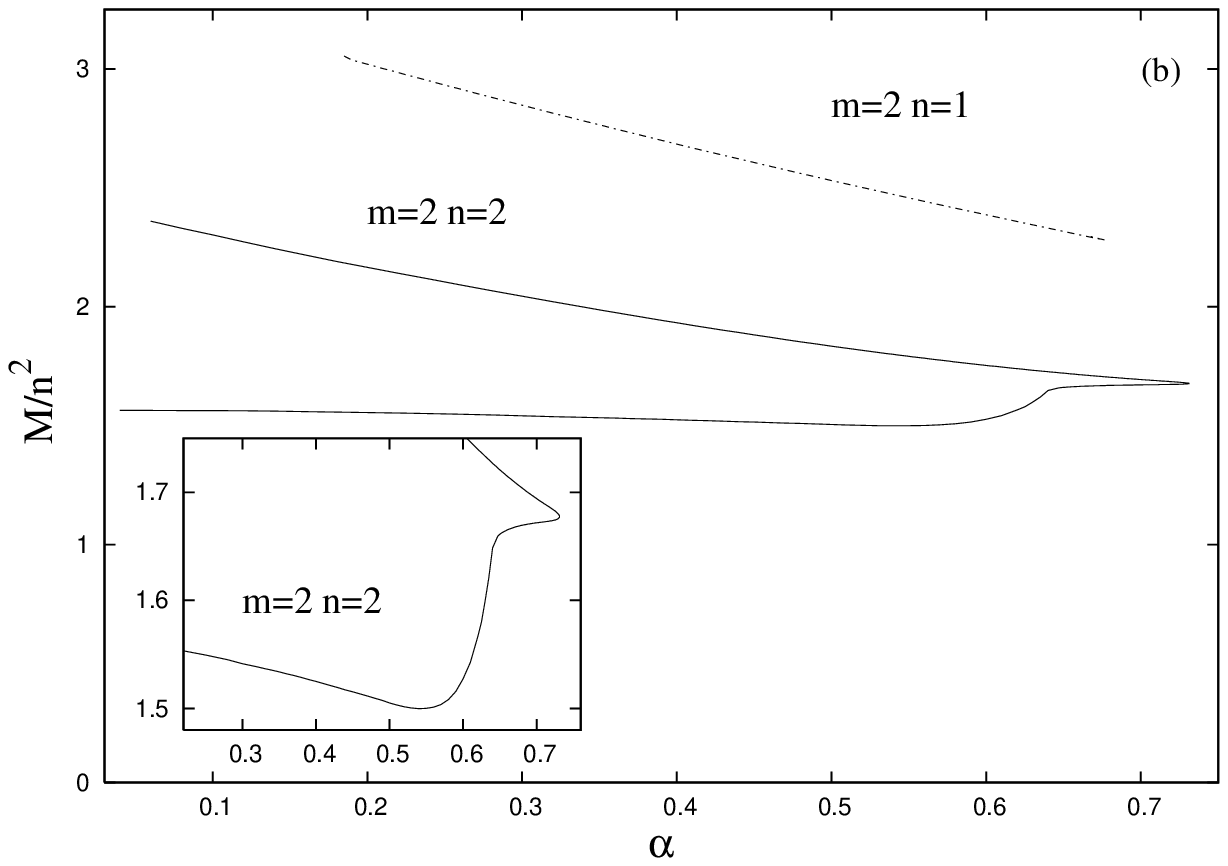}
}}}
\caption{{\small The mass $M$ divided by $n^2$
is shown as a function of $\alpha$ for globally regular EYM solutions:
(a) for $m=1$, and (b) for $m=2$. 
}}
\end{figure}

%%%%%%%%%%%%%%%%%%%%%%%%%%%%%%%%%%%%
For any set of boundary conditions, we have found that the numerical iteration 
fails to converge for $\tau_2=0$. 
Thus, similar to the spherically symmetric case, no reasonable EYM-$p=1$ 
solutions with bi-azimuthal symmetry
is likely to exist. This agrees with the physical intuition based on a 
heuristic Derick-type scaling argument
(although a rigorous proof exists for the spherically symmetric limit 
only \cite{Volkov:2001tb,Okuyama:2002mh}).
It is the $p=2$ YM term, scaling as $L^{-8}$, which enables the existence 
of configurations with finite
mass and well defined asymptotics. 

As in the spherically symmetric case~\cite{Brihaye:2002jg},
dimensionless quantities in this model are obtained by rescaling the radial coordinate
$r \to (\tau_2/\tau_1)^{1/4} r$. This reveals the existence of 
one fundamental parameter which gives the strength of the gravitational interaction
%\begin{eqnarray}
$\alpha^2= \tau_1^{3/2} (16 \pi G/\tau_2^{1/2})$.
%\end{eqnarray}
Thus without loss of generality, one can fix the values of $\tau_1$ and $\tau_2$
to some arbitrary positive values and 
construct the solutions in terms of $\alpha$.
We use this property to set
in the numerical
computation
$\tau_1=\tau_2=1$  for $m=1$ solutions and 
$\tau_1=1,~\tau_2=1/3$  for $m=2$ configurations.

For any set $(m,n)$, the limit $\alpha \to 0$ can be approached in two ways
and two different branches of solutions may exist. The first limit corresponds to a pure $p=1$
YM theory in a flat background ($i.e.$ no gravity and no  $p=2$ YM terms), the solutions here replicating the
(multi-)instantons and composite instanton-antiinstanton bound states discussed in \cite{Radu:2006gg}.
The other possibility corresponds to a finite value of $G$ as $\tau_1 \to 0$. Thus, the 
second limiting configuration is a solution of the truncated system consisting of $p=2$ YM interacting
with gravity, with no $p=1$ YM term.
 
%%%%%%%%%%%%%%%%%%%%%%%%%%%%%%%%%%%%%%%%%%%%%%%%%%%%%%%%%%%%%%%%%%
\subsection{Particle-like solutions}
%%%%%%%%%%%%%%%%%%%%%%%%%%%%%%%%%%%%%%%%%%%%%%%%%%%%%%%%%%%%%%%%%%
%%%%%%%%%%%%%%%%%%%%%%%%%%%%%%%%%%%%%%%%%%%%%%%%%%%%%%%%%%%%%%%%%%
\subsubsection{$m=1$ configurations}
%%%%%%%%%%%%%%%%%%%%%%%%%%%%%%%%%%%%%%%%%%%%%%%%%%%%%%%%%%%%%%%%%%
The $m=1$ configurations carry a topological charge $n^2$ and describe (multi-)solitons.
The $n=1$ spherically symmetric case was discussed in \cite{Brihaye:2002jg} 
in a Schwarzschild coordinate system.
We repeated the numerical analysis of \cite{Brihaye:2002jg} 
using the isotropic coordinate system (\ref{metric}).
In the spherically symmetric limit only two of the functions 
in \re{metric} are independent, $f$ and
$s=l=p$. The dominant term at the gravity decoupling limit $\al\to 0$ is 
the $F(2)$ term, the YM solution being the
well known BPST instanton~\cite{Belavin:1975fg}. When $\alpha$ increases, 
these solutions get deformed by gravity 
and the mass $M$ decreases (see Figure 1a). At the same time, 
the values of the metric functions $f$ and $s$
at the origin decrease, as indicated in Figure 2.
This branch of solutions exists up to a maximal 
value $\alpha_{\rm max}$ of the parameter $\alpha $.
Another branch of solutions is found on the interval $\alpha \in [\alpha_{cr(1)} , \alpha_{\rm max} ]$.
%with $\alpha_{cr(1)} \approx $.
On this second branch of solutions, both $f(0)$ and $s(0)$ 
continue to decrease but stay finite.
However, a third branch of solutions exists for $\alpha \in [\alpha_{cr(1)} , \alpha_{cr(2)}]$ ,
on which the two quantities decrease further. 
A fourth branch of solutions has also been
found, with a corresponding $\alpha_{cr(3)}$ close to $\alpha_{cr(2)}$.  
Along this succession of branches, the values of the metric functions $f$ and $s$
at the origin continue to decrease. 

On the other hand, the mass parameters do not increase significantly along these secondary branches. 
This behaviour with respect to the parameter $\al$ is the same as that which was found
in \cite{Brihaye:2002jg}, for the metric function $\si(r)$ at $r=0$. An analytic explanation of these
results was given in \cite{Breitenlohner:2005hx}, where the observed oscillatory 
behaviour of these functions at
$r=0$ was characterised as a {\it conical fixed point}.  

The $n>1$ non-spherically symmetric solutions are constructed by starting with the known spherically 
symmetric configuration and increasing the winding number $n$ in small steps.
The iterations converge, and repeating the procedure one obtains
in this way solutions for arbitrary $n$. The physical values of $n$ are integers. We have studied $m=1$
solutions with $n=2,3$. As expected, the general features of the spherically symmetric
solutions are the same for all $n>1$ multi-solitons. 
Like for the Yang--Mills dilaton (YMd) model discussed in \cite{Radu:2006qf},
%
%%%%%%%%%%%%%%%%%%%%%%%%%%%%%%%%%%%%
\begin{figure}[h!]
\parbox{\textwidth}
{\centerline{
\mbox{
\epsfysize=15.0cm
\includegraphics[width=82mm,angle=0,keepaspectratio]{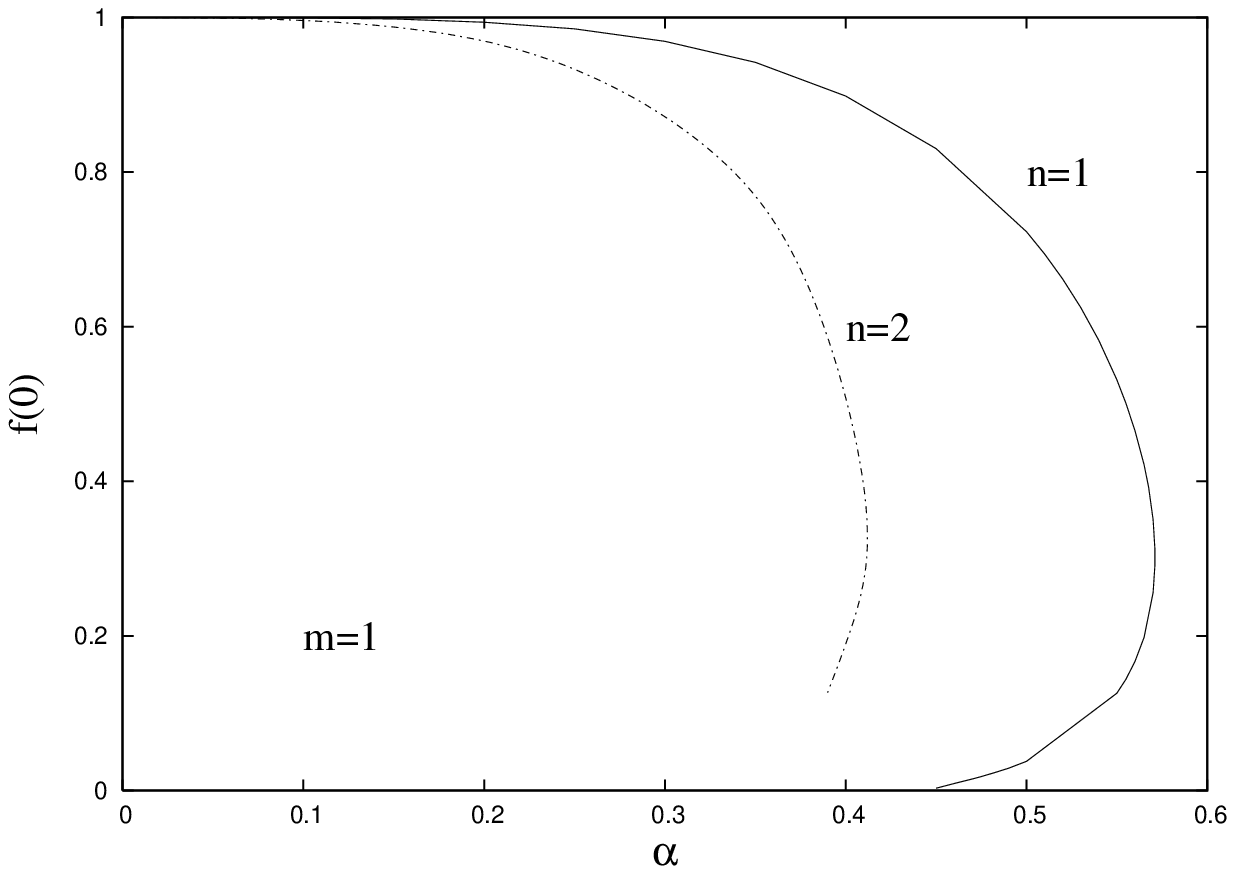}
\includegraphics[width=82mm,angle=0,keepaspectratio]{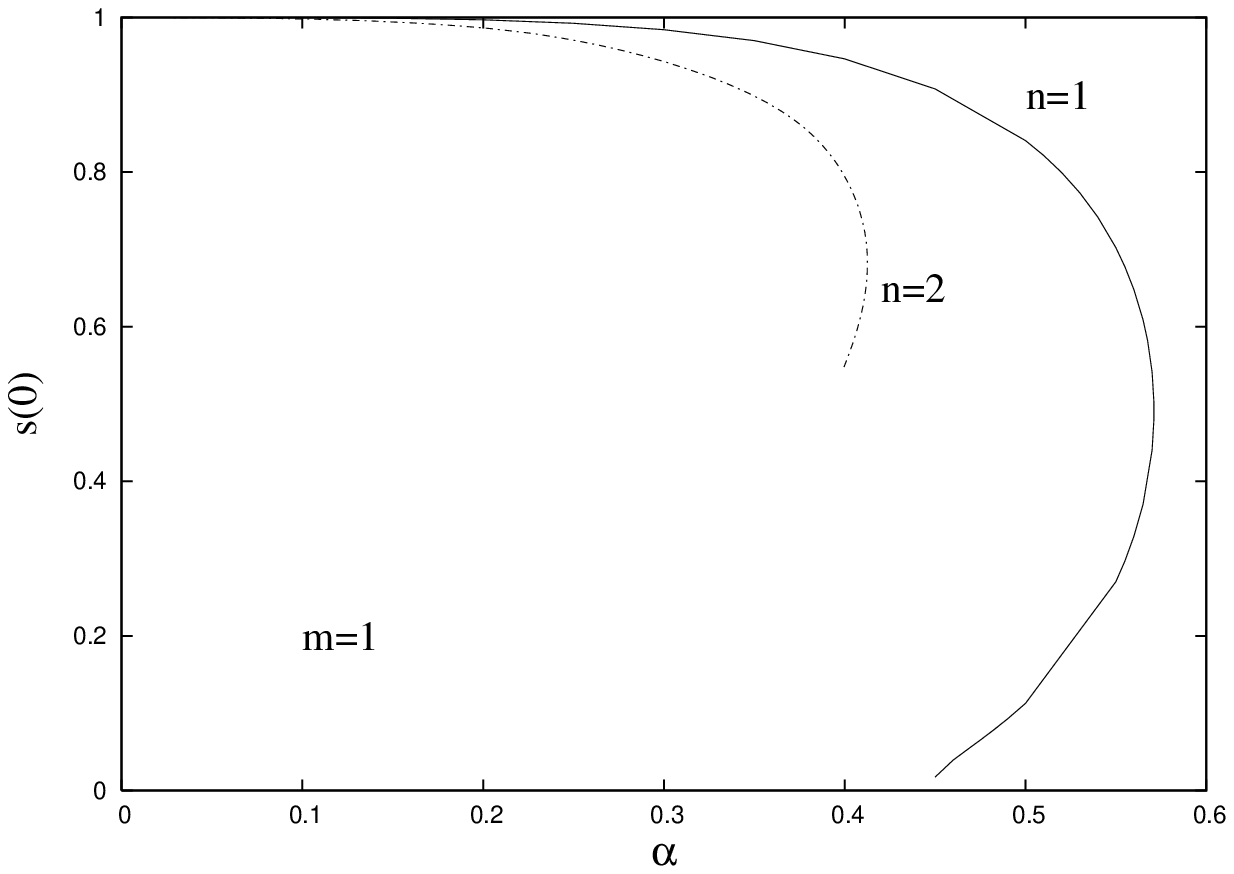}
}}}
\caption{{\small The  value at the origin of the metric functions 
$f$ and $s$ are shown as a function 
of $\alpha$ for  $m=1$ particle-like solutions with $n=1,~2$.}}
\end{figure}
%%%%%%%%%%%%%%%%%%%%%%%%%%%%%%%%%%%%
 when $\alpha$ is increased
from zero, a branch of gravitating solutions with winding number $n$ emerges
smoothly from the corresponding $F(2)$ flat space multi-instanton solution.

This branch extends up to a maximal value $\alpha_{max}(n)$ of 
the coupling constant $\alpha$, beyond which 
the numerical iteration fails to converge. The value of 
$\alpha_{max}(n>1)$ is smaller than the corresponding
value in the spherically symmetric case. For example, we find numerically
$\alpha_{max} (n=2) \approx 0.412$ while the corresponding value 
for $n=1$ is $\alpha_{max} \approx 0.571$. 
For all values $n\geq 1$ we considered, the limiting solutions
 at $\alpha_{max}(n)$ has no special features.
A secondary branch, extending backward in $\alpha$ emerges at $\alpha_{max}(n)$.
However, the numerical accuracy deteriorates drastically for the 
secondary branch of solutions around some
critical value $\alpha_{cr} \sim 0.38$. Our numerical results in this case are less conclusive,
the properties of these configurations requiring further work.
We notice, however, that the value at the origin of all metric functions decreases 
along these branches, as seen in Figure 2~\footnote{Note that the
values at the origin of all metric functions exhibited in this paper
correspond to $f(r=0,\ta=0)$,~$s(r=0,\ta=0)$. This restriction is reasonable since for all solutions
with bi-azimuthal symmetry that we have found, the metric functions at $r=0$ 
present almost no dependence on
the angle $\theta$.}. 
We expect that the oscillatory pattern of $g_{tt}(0)$ arising 
from the {\it conical fixed point}
observed  for the spherically symmetric $m=1,~n=1$ solutions, will also be discovered for the
$n>1$ solutions here. However, the construction of the secondary
branches of solutions is a difficult numerical problem beyond the scope of the present work.

In all cases we have studied, the metric functions $f,~l,~p,~s$ are completely regular
and show no sign of an apparent horizon, while $l$ and $p$ have rather similar shapes.
The angular dependence of the metric functions is rather small, although it increases somewhat with $n$.
The gauge functions $a_r,a_\theta,\chi^A,\xi^A$ look very similar to those of the 
YMd solutions presented in \cite{Radu:2006qf}.
Both $|\chi|= ((\chi^1)^2+(\chi^2))^{1/2}$ and $|\xi|=((\xi^1)^2+(\xi^2))^{1/2}$ 
possess one node on the $\theta=0$ and $\theta=\pi/2$ axis, respectively.
The positions of these nodes move inward along the branches.

It is also interesting to note that for the $m=1$ solutions, the mass per unit 
charge of the gravitating
multisoliton solutions is lower than the mass of a single particle, see Figure 1a.
Thus these multisolitons are gravitationally bound states.
This case resembles the situation found for $d=4$ gravitating EYMH monopoles
with a vanishing or small Higgs selfcoupling \cite{HKK}.

%%%%%%%%%%%%%%%%%%%%%%%%%%%%%%%%%%%%
\begin{figure}[h!]
\parbox{\textwidth}
{\centerline{
\mbox{
\epsfysize=15.0cm
\includegraphics[width=82mm,angle=0,keepaspectratio]{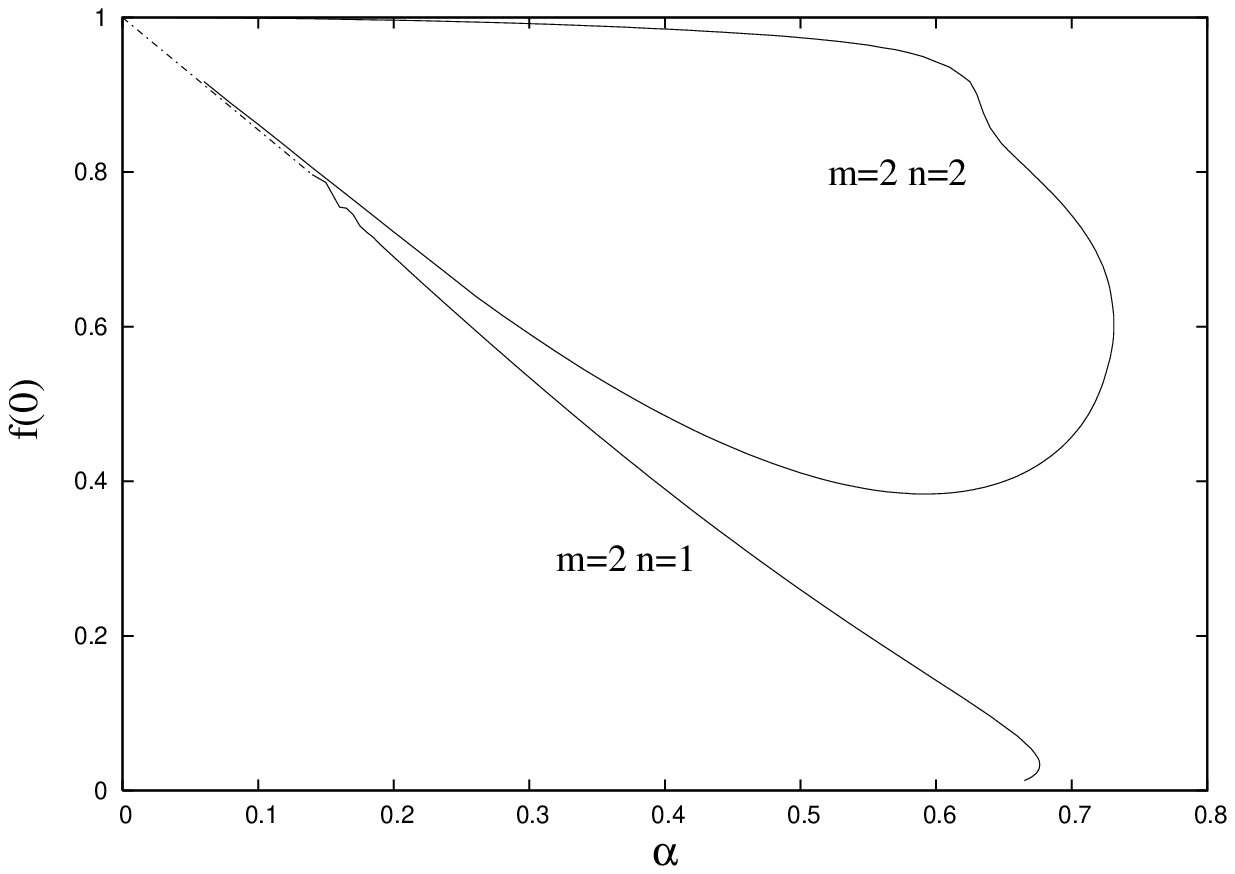}
\includegraphics[width=82mm,angle=0,keepaspectratio]{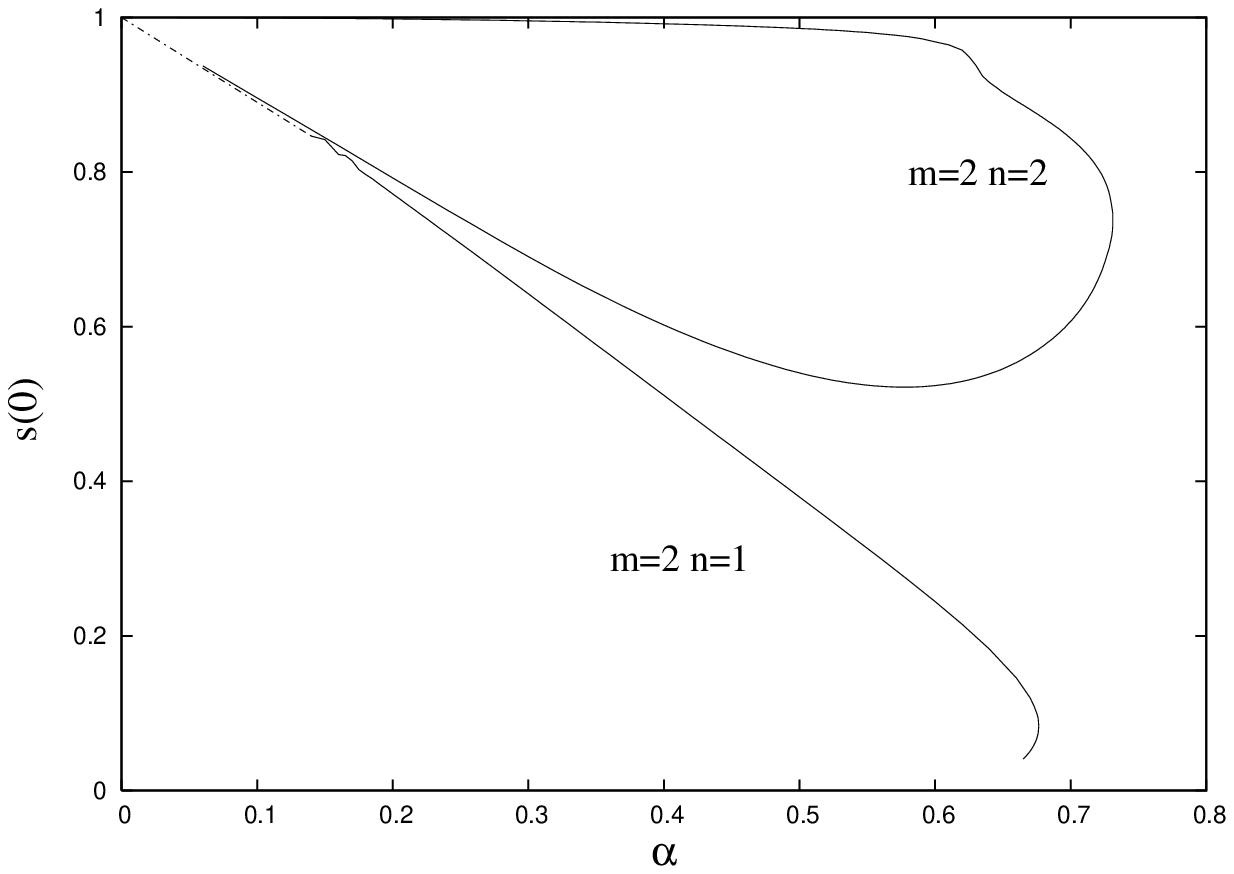}
}}}
\caption{{\small The same as Figure 2 for  $m=2$ particle-like solutions with $n=1,~2$.}}
\end{figure}
%%%%%%%%%%%%%%%%%%%%%%%%%%%%%%%%%%%%%
%%%%%%%%%%%%%%%%%%%%%%%%%%%%%%%%%%%%%%%%%%%%%%%%%%%%%%%%%%%%%%%%%%
\subsubsection{$m=2$ configurations}
%%%%%%%%%%%%%%%%%%%%%%%%%%%%%%%%%%%%%%%%%%%%%%%%%%%%%%%%%%%%%%%%%%
The $m=2$ configurations reside in the topologically trivial
sector. These solutions can be thought of as composite systems
consisting of two components which are pseudoparticles of
Chern-Pontryagin topological charges $\pm n^2$. This type of
solutions have no spherically symmetric limit. The position of
each constituent can be identified according to the location of
the maxima of the energy density. Also, the structure and location
of the nodes of the (effective Higgs) scalar fields nicely reveal the
evolution and the types of the solutions present at the respective
values of the gravitational strength.

As in the case of the $m=1$ configurations, coupling with
gravity yields various branches of gravitating solutions which,
however, have different limits depending on the values of the topological
charge $n^2$ of the constituents. Also, their behaviours as functions
of the gravitational coupling $\alpha$ differ from those with
$m=1$ presented above.

\subsubsection*{$n=1$}
There is a certain similarity between the properties of
the $4+1$ dimensional YMd model studied in \cite{Radu:2006qf}, and
the model under consideration here. As in the former case, we find that
in the limit $\alpha\to 0$ resulting from $G \to 0$, no solution with $n=1$
exists, {\it i.e.} that in the gravity decoupling limit no such solution exists.
On the other hand, we know from the work of \cite{Brihaye:2002jg} that in the flat space
limit the EYM solution of this model reduces to the BPST instanton~\cite{Belavin:1975fg}
of the $p=1$ (usual) YM model, so that in this limit the $p=1$ YM term dominates over
the $p=2$ term. Thus the nonexistence of a $m=2\ ,\ n=1$ solution here in the
gravity decoupling limit implies that there should exist no such solution in the
$4+0$ dimensional $p=1$ YM model on flat space. This is precisely what was found
in \cite{Radu:2006gg}.

In the other limit of $\alpha \to 0$ however, when both $\tau_1 \to 0$ and the gravitational
coupling $G$ remain finite, such solutions exist. It turns out that in this limit, it
is the $p=2$ term which dominates over the $p=1$ YM term. 
The characteristic feature of this this configuration is that both nodes of the effective Higgs fields
$|\chi| $ and $|\xi|$ merge on the $\theta = \pi/4$ hypersurface. From this limiting configuration, a branch
evolves as $\alpha$ increases. Along this branch the nodes move towards the symmetry axes, $\rho$ and $\sigma$,
respectively (with $\rho=r\sin \theta$, $\sigma=\rho \cos \theta$), forming two identical vortex rings whose
radii slowly decrease, while the separation of both rings from the origin also decreases.
The evolution of the solution along this branch can be associated with the increase of the coupling
$\tau_1$, while $\tau_2$ and the gravitational coupling $G$ remain fixed. This
reproduces the corresponding pattern in the YMd
system~\cite{Radu:2006qf}. Note that there is a difference between the evolutions of the configurations
we are considering here in this $4+1$ dimensional theory, and the behaviour of the gravitating multimonopoles
or the monopole-antimonopole solutions of the gravitating YMH system $3+1$ theory~\cite{HKK,KKS}.
Although the latter also feature different branches, the evolution along those branches is
usually associated with the increasing of the gravitational coupling $G$ on the lower
mass branch, and, the decreasing of the VEV of the Higgs field on the upper mass branch. More
importantly, the $m=2\ ,\ n=1$ solution in that case does have a gravity decoupling limit.
%
%
%%%%%%%%%%%%%%%%%%%%%%%%
\begin{figure}[h!]
\parbox{\textwidth}
{\centerline{
\mbox{
\epsfysize=10.0cm
\includegraphics[width=92mm,angle=0,keepaspectratio]{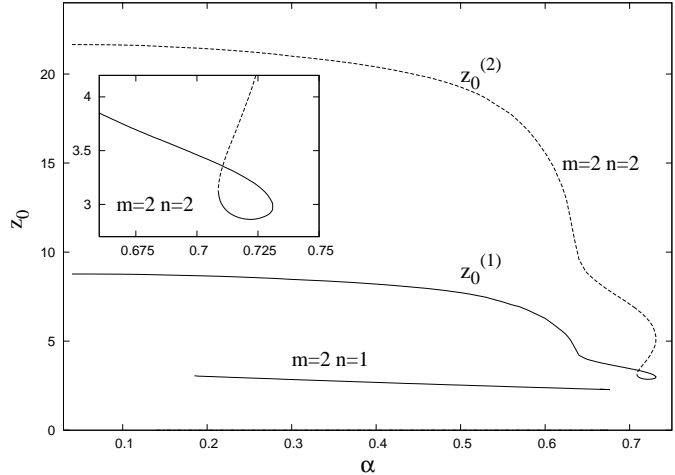} 
}}}
\caption{{\small The   position of nodes $(z_0^{(1)},~z_0^{(2)})$ 
is  presented   for $m=2$ particle-like solutions with $n=1,~2$.}}
\end{figure}
%%%%%%%%%%%%%%%%%%%%%%%%
Thus, the gravitating solutions of the $3+1$ YMH theory usually are linked to flat
space configurations, while the solutions discussed here clearly
do not have a flat space limit.

On the $p=2$ branch (where the $p=2$ term ${\cal F}_{MNRS}^2$ dominates) of 
five dimensional EYM $m=2\ , n=1$ solutions, the gauge functions  $a_r$, $a_\theta$ as well
the metric functions $f$ and $s$ are almost $\theta$-independent,
whereas the metric functions $l$
and $p$ possess reflection symmetry with respect to $\theta = \pi/4$ axis. 
As seen in Figures 1 and 2,
the mass of the gravitating solutions on this branch decreases, as well as the values at 
the origin of the metric functions.

At the critical value $\alpha \simeq 0.672$, the node structure
of the configuration changes and both vortex rings shrink to zero
size, two isolated nodes appearing on each symmetry axis.
This transition means that the $p=1$ term ${\cal F}_{MN}^2$
becomes dominant. This secondary branch has a small extension in $\alpha$ up to the
maximal value $\alpha_{max}\simeq0.6765$, beyond which we could not find
regular gravitating solutions. We found instead that this branch
merges here with the second, $p=1$ branch, which evolves backwards in $\alpha$ as the value of
the metric function $f(0)$ continues to decrease. 

The evolution along this short branch can be associated with the decrease of the coupling constant
$\tau_2$ relative to $\tau_1$, 
as the gravitational coupling $G$ remains fixed. For this branch the
relative distance between the nodes increases, 
one lump slowly moving towards the origin and the other one moving in the
opposite direction. This branch persists up to a value  of the coupling constant $\alpha_{cr} \simeq 0.6665$,
where a critical solution is approached. Due to severe numerical difficulties encountered here,
we could not clarify the properties of this critical solution further. 
As $\alpha\to \alpha_{cr}$,  
the metric function $f(0)$ takes a very small value, $f(0)\simeq 10^{-3}$, 
while $s(0)$ remains one order of magnitude larger (see Figure 3).
At the same time, the Lagrangian density and the mass of the configuration 
remain finite at that point.
The critical behaviour observed here resembles the case of the gravitating $4+1$ EYM vortices
in the model consisting only of the $p=1$ YM term \cite{Volkov:2001tb}. 
It is tempting to speculate that, similar to case in \cite{Volkov:2001tb}, 
the solution splits into two parts:
a non-singular interior region with a special geometry  (so-called throat) 
and an exterior asymptotically flat
region where two pseudoparticles are located. However, another parametrisation 
of the metric, differing
from (\ref{metric}) (and possibly even a different numerical approach) appears 
to be necessary to clarify these aspects.
 
%%%%%%%%%%%%%%%%%%%%%%%%%%%%%%%%%%%%%%%%%%%%%%%%%%%%%
\subsubsection*{$n=2$}
%%%%%%%%%%%%%%%%%%%%%%%%%%%%%%%%%%%%%%%%%%%%%%%%%%%%%
This configuration also resides in the topologically trivial sector 
and can be considered as consisting of two pseudoparticles
of charges $\pm 2^2$. In this case the interaction between 
the non Abelian matter fields becomes stronger than in the case
of $\pm 1$ constituents, resulting in a different pattern 
of possible branches of solutions.
Indeed, as in the case of the $4+1$ dimensional YMd system 
\cite{Radu:2006qf}, we observe two different branches of
gravitating solutions, both linked to the $\alpha\to 0$ limit.  
%
%%%%%%%%%%%%%%%%%%%%%%%%
\begin{figure}[h!]
\parbox{\textwidth}
{\centerline{
\mbox{
\epsfysize=10.0cm
\includegraphics[width=92mm,angle=0,keepaspectratio]{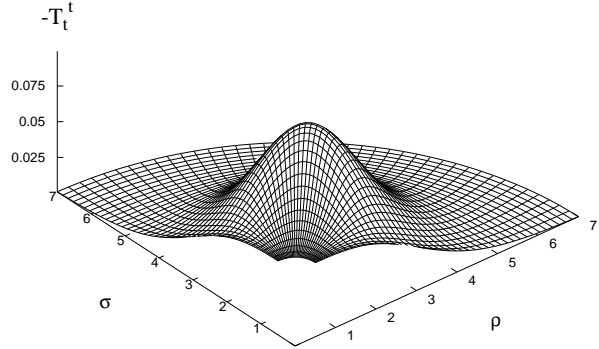} 
}}}
\caption{{\small
The energy density $\epsilon=-T_t^t$ is shown as a function of the coordinates
$\rho=r\sin \theta$, $\sigma=\rho \cos \theta$ for a $m=1,~n=2$ 
EYM black hole solutions with
$\alpha=0.2$, $r_h=0.5$. }}
\end{figure}
%%%%%%%%%%%%%%%%%%%%%%%%
%
The lower branch, on which the $p=1$ YM term dominates, emerges from the corresponding flat
space solution of the pure YM theory with vanishing $p=2$ term.
Varying $\al$ along this branch is associated with the decrease of $\tau_1$, at fixed $\tau_2$ and
fixed gravitational coupling $G$. 

For small values of $\alpha$ the corresponding $m=2\ , n=2$
solutions possess two (double) nodes of the fields $|\chi| $ and
$|\xi|$ on the $\rho$ and $\sigma$ symmetry axes, respectively.
The locations of nodes correspond to the locations of the two
individual constituents and the action density distribution
possesses two distinct maxima on the $\theta = \pi/4$ axis. As
$\alpha$ increases the mass of the solution increases and both
pseudoparticles move from spatial infinity towards the origin. 
For values of $\alpha$ smaller than $\alpha_{cr} \simeq 0.635$ along this branch, 
the energy of interaction between the individual pseudoparticles is relatively small 
and both constituents remain individual. We observe that, as the coupling constant 
approaches this critical value from below, the energy of interaction rapidly increases and 
both pseudoparticles form a bound state, as seen from Figures 3, 4.

This branch extends further up to a maximal value $\alpha_{max} \simeq 0.7265$ 
where it bifurcates with an upper $p=2$ branch which
extends all the way back to $\alpha = 0$. Varying $\al$ along this branch 
is associated with the increase of $\tau_2$ relative to
$\tau_1$, as gravitational constant $G$ remains fixed.

Along the upper branch, as $\alpha$ slightly decreases below
$\alpha_{max}$, the inner node inverts direction of its movement
toward the outer node which still moves inwards. Thus, both nodes
on the symmetry axis rapidly approach each other and merge forming
a two vortex ring solution at $\alpha \simeq 0.708$. The action
density then has a single maximum on $\theta = \pi/4$ axis. As
$\alpha$ decreases further both nodes move away from the symmetry
axis and their positions do not coincide with the location of the
maximum of the action density. Further decreasing $\alpha$ results
in the increase of the radii of the two rings around the symmetry
axis, and in the limit  $\alpha\to 0$ the rings touch each other
on the $\theta = \pi/4$ hyperplane.

%%%%%%%%%%%%%%%%%%%%%%%%%%%%%%%%%%%%%%%%%%%%%%%%%%%%%%%%%%%%%%%%%%
\subsection{Black hole solutions}
%%%%%%%%%%%%%%%%%%%%%%%%%%%%%%%%%%%%%%%%%%%%%%%%%%%%%%%%%%%%%%%%%%
According to the standard arguments, one can expect
black hole generalisations of the regular configurations to exist at
least for small values of the horizon radius $r_h$.
This is confirmed by the numerical analysis for $m=1,~n=2$.
Several  black hole solutions with  $m=2,~n=1$ have been also constructed, with a lower
numerical accuracy, however. 

As discussed in \cite{Brihaye:2002jg} spherically symmetric  $m=1,\ n=1$ black hole counterparts 
exist for any regular solution with the same amount of symmetry. 
Starting for a given $\alpha_0<\alpha_{\rm max}$ from a $r_h=0$ first branch
regular solution,  one finds a branch of black hole solutions
extending up to a maximal value of the event horizon radius $r_h=r_h^{\rm max}$.
When $r_h$ increases, both the mass and the Hawking temperature increase.
The value of $r_{h(max)}$ depends on $\alpha$. The Hawking temperature decreases on this branch,
while the mass parameter increases; however, the variation of mass is relatively small. 
The corresponding picture for secondary branches is more complicated and will not be
discussed here.

The numerical construction of nonspherically symmetric black hole solutions
appears to be more difficult than in the globally regular case. However, our numerical results
indicate that the $m=1,~n>1$ black hole solutions with bi-azimuthal symmetry
 follow this general pattern.
First, black hole solutions seem to exist for all values of $\alpha$ or 
which regular configurations could be
constructed (here we restrict again to first branch solutions). Also, 
it appears that black hole 
solutions exist only for a limited region of the $(r_h,\alpha)$ space. 
However, for a given value of $\alpha$,
it is very difficult to find an accurate value of $r_h^{\rm max}$. 
An approach to this problem with a different method
appears to be necessary.

These solutions possess a regular deformed $S^3$ horizon. The energy density has a pronounced 
angle-dependence, with a maximum on the $\theta=\pi/2$ hypersurface. 
Figure 5 shows a three dimensional plot of the 
energy density of a $m=1\ ,\ n=2$ black hole with $\alpha=0.2,~r_h=0.5$
as a function of the coordinates $\rho=r \sin \theta,~\sigma=r\cos \theta$.
With increasing the winding number $n$, the absolute maximum of the energy density 
residing on the $\rho=\sigma$ axis, shifts inward. The metric and gauge functions possess
a nontrivial angular dependence at the horizon. 

Outside their event horizon, these black holes possess nontrivial non Abelian fields.
Therefore they represent a further counterexample to the $d=5$ no-hair conjecture.
Also, these bi-azimuthally symmetric black holes clearly show that the higher dimensional static
black hole solutions need not be spherically symmetric.

%%%%%%%%%%%%%%%%%%%%%%%%%%%%%%%%%%%%%%%%%%%%%%%%%%%%%%%%%%%%%%%%%%%%%%%%%%%%%%
\section{Conclusions}
%%%%%%%%%%%%%%%%%%%%%%%%%%%%%%%%%%%%%%%%%%%%%%%%%%%%%%%%%%%%%%%%%%%%%%%%%%%%%
Motivated by the recent interest in gravitating solutions in higher dimensional
spacetime, we have studied static, bi-azimuthally symmetric solutions with non Abelian fields
in  $d=4+1$
%five 
spacetime dimensions. Our solutions are akin to the static, axially symmetric
EYM configurations in $d=4$, studied exhaustively in \cite{Kleihaus:1997mn},
\cite{Ibadov:2005rb,Ibadov:2004rt}. Our choice
of bi-azimuthal symmetry is motivated by our desire to reduce the boundary value problem to a
two dimensional one. An alternative symmetry imposition resulting in a two dimensional
residual system would be imposition of $SO(3)$ spherical symmetry in the $3$ dimensional
spacelike dimensions like in \cite{Witten:1976ck}. We have eschewed this alternative for
purely technical reasons (see footnote $^1$).

The regular and black hole solutions presented are natural generalisations 
of the known~\cite{Brihaye:2002jg}
$d=5$ EYM spherically symmetric globally regular and black hole solutions. 
Like the former they
are asymptotically flat, finite mass solutions, that describe nontrivial gravitating
magnetic gauge field configurations. Our $d=5$ EYM configurations are the first $d\ge 5$
dimensional static solutions in the literature, which are not spherically symmetric.

In the case of particle like solutions, which we have studied much more intensively than
their black hole counterparts, their dependence on the effective gravity coupling $\al$ is
analysed numerically in some detail. By and large this is qualitatively very similar to
that for the YMd solutions~\cite{Radu:2006qf} in $4+1$ dimensions, except that here
we have four metric functions to keep track of, as opposed to the single dilaton field in
the previous case~\cite{Radu:2006qf}. We have studied regular solutions with
$m=2\ ,\ n=1$ and $m=2\ ,\ n=2$ in detail, numerically.

Just as in the YMd case, here too there exists a $m=2\ ,n=1$ solution 
on the branch where the $p=2$ YM term dominates,
while on the other branch, where the $p=1$ YM term dominates, 
such a solution is absent. As it turns out the
$p=1$ YM term dominates in the gravity decoupling limit, which 
is consistent with our knowledge that this model
in $4+0$ dimensions does not support~\cite{Radu:2006gg} a $m=2\ ,n=1$ solution.

Another qualitative feature of $5$ dimensional EYM solutions that is 
confirmed here is the occurrence of
a {\it conical} singular behaviour with respect to the dependence of 
the metric functions on $\al$. 
This features
the oscillatory picture first discovered for the $m=1,\ n=1$ spherically 
symmetric solutions in \cite{Brihaye:2002jg}
and analysed in \cite{Breitenlohner:2005hx}, which are found also here for 
the $m=1\ ,\ n>1$ case.
 
As compared to the $d=4$ case~\cite{Ibadov:2005rb,Ibadov:2004rt}, we expect 
the existence of a much richer
set of nonspherically symmetric EYM solutions in $d=5$. 
The configurations studied here represent only the simplest,
asymptotically flat type of $d=5$ nonspherically symmetric gravitating nonabelian solutions.
For example, it is known that $d=5$ Einstein gravity coupled to 
Abelian fields presents black
ring~\cite{Elvang:2003yy} solutions. These solutions 
have an horizon topology $S^2 \times S^1$ and approach at
infinity the flat ${\cal M}^{5}$ background, as is the case with our solutions. 
It would be interesting to
construct non Abelian versions of the $U(1)$ black ring solutions. 
A black ring can be constructed in a heuristic
way by taking a black string, bending the extra dimension and spinning it
 along the circle direction just enough
so that the gravitational attraction is balanced by the centrifugal force. 
In this framework
the (putative) nonabelian black ring would behave locally as a boosted black string, e.g. that in
\cite{Brihaye:2005tx}, with very similar charges and fields. 
The numerical work involved in the construction of
non spherically symmetric higher dimensional EYM solutions is, however, 
a considerably challenging task.

%%%%%%%%%%%%%%%%%%%%%%%%%%%%%%%%%%%%%%%%%%%%%%%%%%%%%%%%%%%%%%%%%%%
\bigskip
\noindent
{\bf\large Acknowledgements} \\
We thank  Mikhail Volkov for helpful comments and discussions.
This work was carried out in the framework of Science Foundation 
Ireland
(SFI) Research Frontiers Programme (RFP) project RFP07/FPHY330.
 The collaboration with Ya. Shnir
was supported by a Research Enhancement Grant from the Office of
the Dean of Research and Postgraduate Studies of the NUI Maynooth.\begin{small}

\end{small}


\begin{thebibliography}{99}
%%%%%%%%%%%%%%%%%%%%%%%%%%%%%%%%%%%%%%%%%%%%%%%%%%%%%%%%%%%%%%%%%%%%%%%%%%%%%%
%\cite{Volkov:1998cc}
\bibitem{Volkov:1998cc}
  M.~S.~Volkov and D.~V.~Gal'tsov,
  %``Gravitating non-Abelian solitons and black holes with Yang-Mills  fields,''
  Phys.\ Rept.\  {\bf 319} (1999) 1
  [arXiv:hep-th/9810070].
  %%CITATION = HEP-TH 9810070;%%  
%%%%%%%%%%%%%%%%%%%%%%%%%%%%%%%%%%%%%%%%%%%%%%%%%%%%%%%%%%%%%%%%%%%%%%%%%%%%%%
%\cite{Volkov:2006xt}
\bibitem{Volkov:2006xt}
  M.~S.~Volkov,
  %``Gravitating non-Abelian solitons and hairy black holes in higher
  %dimensions,''
  arXiv:hep-th/0612219.
  %%CITATION = HEP-TH/0612219;%
%%%%%%%%%%%%%%%%%%%%%%%%%%%%%%%%%%%%%%%%%%%%%%%%%%%%%%%%%%%%%%%%%%%%%%%%%%%%%%
%\cite{Tangherlini:1963bw}
\bibitem{Tangherlini:1963bw}
  F.~R.~Tangherlini,
  %``Schwarzschild field in n dimensions and the dimensionality of space
  %problem,''
  Nuovo Cim.\  {\bf 27} (1963) 636.
  %%CITATION = NUCIA,27,636;%%
%%%%%%%%%%%%%%%%%%%%%%%%%%%%%%%%%%%%%%%%%%%%%%%%%%%%%%%%%%%%%%%%%%%%%%%%%%%%%%
%\cite{Myers:1986un}
\bibitem{Myers:1986un}
  R.~C.~Myers and M.~J.~Perry,
  %``Black Holes In Higher Dimensional Space-Times,''
  Annals Phys.\  {\bf 172} (1986) 304.
  %%CITATION = APNYA,172,304;%%
%%%%%%%%%%%%%%%%%%%%%%%%%%%%%%%%%%%%%%%%%%%%%%%%%%%%%%%%%%%%%%%%%%%%%%%%%%%%%%
%\cite{Emparan:2001wn}
\bibitem{Emparan:2001wn}
  R.~Emparan and H.~S.~Reall,
  %``A rotating black ring in five dimensions,''
  Phys.\ Rev.\ Lett.\  {\bf 88} (2002) 101101
  [arXiv:hep-th/0110260].
  %%CITATION = PRLTA,88,101101;%%
%%%%%%%%%%%%%%%%%%%%%%%%%%%%%%%%%%%%%%%%%%%%%%%%%%%%%%%%%%%%%%%%%%%%%%%%%%%%%%
%\cite{Elvang:2007rd}
\bibitem{Elvang:2007rd}
  H.~Elvang and P.~Figueras,
  %``Black Saturn,''
  arXiv:hep-th/0701035.
  %%CITATION = HEP-TH/0701035;%%
%%%%%%%%%%%%%%%%%%%%%%%%%%%%%%%%%%%%%%%%%%%%%%%%%%%%%%%%%%%%%%%%%%%%%%%%%%%%%%
%\cite{Horowitz:1991cd}
\bibitem{Horowitz:1991cd}
  G.~T.~Horowitz and A.~Strominger,
  %``Black strings and P-branes,''
  Nucl.\ Phys.\  B {\bf 360} (1991) 197.
  %%CITATION = NUPHA,B360,197;%% 
%%%%%%%%%%%%%%%%%%%%%%%%%%%%%%%%%%%%%%%%%%%%%%%%%%%%%%%%%%%%%%%%%%%%%%%%%%%%%%
%\cite{Bartnik:1988am}
\bibitem{Bartnik:1988am}
  R.~Bartnik and J.~McKinnon,
  %``Particle - Like Solutions Of The Einstein Yang-Mills 
  Phys.\ Rev.\ Lett.\  {\bf 61} (1988) 141.
  %%CITATION = PRLTA,61,141;%%
%%%%%%%%%%%%%%%%%%%%%%%%%%%%%%%%%%%%%%%%%%%%%%%%%%%%%%%%%%%%%%%%%%%%%%%%%%%%%
%\cite{Brihaye:2002hr}
\bibitem{Brihaye:2002hr}
  Y.~Brihaye, A.~Chakrabarti and D.~H.~Tchrakian,
  %``Particle-like solutions to higher order curvature Einstein-Yang-Mills
  %systems in d dimensions,''
  Class.\ Quant.\ Grav.\  {\bf 20} (2003) 2765
  [arXiv:hep-th/0202141].
  %%CITATION = HEP-TH 0202141;%
%%%%%%%%%%%%%%%%%%%%%%%%%%%%%%%%%%%%%%%%%%%%%%%%%%%%%%%%%%%%%%%%%%%%%%%%%%%%%
%\cite{Brihaye:2002jg}
\bibitem{Brihaye:2002jg}
  Y.~Brihaye, A.~Chakrabarti, B.~Hartmann and D.~H.~Tchrakian,
  %``Higher order curvature generalisations of Bartnick-McKinnon and coloured
  %black hole solutions in d = 5,''
  Phys.\ Lett.\ B {\bf 561} (2003) 161
  [arXiv:hep-th/0212288].
  %%CITATION = HEP-TH 0212288;%%
%\cite{Brihaye:2004gv}
\bibitem{Brihaye:2004gv}
  Y.~Brihaye, E.~Radu and D.~H.~Tchrakian,
  %``Regular and black hole solutions to higher order curvature
  %Einstein-Yang-Mills-Grassmannian systems in 5 dimensions,''
  Int.\ J.\ Mod.\ Phys.\  A {\bf 19} (2004) 5085
  [arXiv:hep-th/0405255].
  %%CITATION = IMPAE,A19,5085;%%
%%%%%%%%%%%%%%%%%%%%%%%%%%%%%%%%%%%%%%%%%%%%%%%%%%%%%%%%%%%%%%%%%%%%%%%%%%%
%%%%%%%%%%%%%%%%%%%%%%%%%%%%%%%%%%%%%%%%%%%%%%%%%%%%%%%%%%%%%%%%%%%%%%%%%%%%%%
%\cite{Breitenlohner:2005hx}
\bibitem{Breitenlohner:2005hx}
  P.~Breitenlohner, D.~Maison and D.~H.~Tchrakian,
  %``Regular solutions to higher order curvature Einstein-Yang-Mills systems  in
  %higher dimensions,''
  Class.\ Quant.\ Grav.\  {\bf 22} (2005) 5201
  [arXiv:gr-qc/0508027].
  %%CITATION = GR-QC 0508027;%%
%%%%%%%%%%%%%%%%%%%%%%%%%%%%%%%%%%%%%%%%%%%%%%%%%%%%%%%%%%%%%%%%%%%%%%%%%%%%
%\cite{Radu:2005mj}
\bibitem{Radu:2005mj}
  E.~Radu and D.~H.~Tchrakian,
  %``No hair conjecture, nonabelian hierarchies and anti-de Sitter  spacetime,''
  Phys.\ Rev.\ D {\bf 73} (2006) 024006
  [arXiv:gr-qc/0508033].
  %%CITATION = GR-QC 0508033;%%
%%%%%%%%%%%%%%%%%%%%%%%%%%%%%%%%%%%%%%%%%%%%%%%%%%%%%%%%%%%%%%%%%%%%%%%%%%%
%\cite{Radu:2006mb}
\bibitem{Radu:2006mb}
  E.~Radu, C.~Stelea and D.~H.~Tchrakian,
  %``Features of gravity-Yang-Mills hierarchies in d-dimensions,''
  Phys.\ Rev.\ D {\bf 73} (2006) 084015
  [arXiv:gr-qc/0601098].
  %%CITATION = GR-QC 0601098;%%
%%%%%%%%%%%%%%%%%%%%%%%%%%%%%%%%%%%%%%%%%%%%%%%%%%%%%%%%%%%%%%%%%%%%%%%%%%%%%%
%\cite{Brihaye:2006xc}
\bibitem{Brihaye:2006xc}
  Y.~Brihaye, E.~Radu and D.~H.~Tchrakian,
  %``Einstein-Yang-Mills solutions in higher dimensional de Sitter spacetime,''
  Phys.\ Rev.\  D {\bf 75} (2007) 024022
  [arXiv:gr-qc/0610087].
  %%CITATION = PHRVA,D75,024022;%%
%%%%%%%%%%%%%%%%%%%%%%%%%%%%%%%%%%%%%%%%%%%%%%%%%%%%%%%%%%%%%%%%%%%%%%%%%%%%%%%
%\cite{Tchrakian:1984gq}
\bibitem{Tchrakian:1984gq}
  D.~H.~Tchrakian,
  %``Spherically Symmetric Gauge Field Configurations With Finite Action In 4
  %P-Dimensions (P = Integer),''
  Phys.\ Lett.\  B {\bf 150} (1985) 360.
  %%CITATION = PHLTA,B150,360;%%
%%%%%%%%%%%%%%%%%%%%%%%%%%%%%%%%%%%%%%%%%%%%%%%%%%%%%%%%%%%%%%%%%%%%%%%%%%%
\bibitem{hier}
D.H. Tchrakian, {\it Yang-Mills hierarchy}, in Differential
Geometric Methods in Theoretical Physics, eds. C.N. Yang, M.L. Ge and
X.W. Zhou, Int. J. Mod. Phys. A (Proc.Suppl.) {\bf 3A} (1993) 584.
%%%%%%%%%%%%%%%%%%%%%%%%%%%%%%%%%%%%%%%%%%%%%%%%%%%%%%%%%%%%%%%%%%%%%%%%%%%%%%%%%%%%%%%%%
%%\cite{Tseytlin}
\bibitem{Tseytlin}
A.A. Tseytlin, {\it Born--Infeld action, supersymmetry and string 
theory},
in {\it Yuri Golfand memorial volume}, ed. M. Shifman, World 
Scientific (2000).
%%%%%%%%%%%%%%%%%%%%%%%%%%%%%%%%%%%%%%%%%%%%%%%%%%%%%%%%%%%%%%%%%%%%%%%%%%%%
%%\cite{BRS}
\bibitem{BRS}
E. Bergshoeff, M. de Roo and A. Sevrin, Fortsch.Phys. 49 (2001) 
433-440;
Nucl.Phys.Proc.Suppl. 102 (2001) 50-55.
%\cite{Bergshoeff:2000cx}
%\bibitem{Bergshoeff:2000cx}
  E.~A.~Bergshoeff, M.~de Roo and A.~Sevrin,
  %``On the supersymmetric non-Abelian Born-Infeld action,''
  Fortsch.\ Phys.\  {\bf 49} (2001) 433
  [Nucl.\ Phys.\ Proc.\ Suppl.\  {\bf 102} (2001) 50]
  [arXiv:hep-th/0011264].
  %%CITATION = HEP-TH 0011264;%%
%%%%%%%%%%%%%%%%%%%%%%%%%%%%%%%%%%%%%%%%%%%%%%%%%%%%%%%%%%%%%%%%%%%%%%%%%%%%%
%%\cite{Cederwall:2001bt}
\bibitem{Cederwall:2001bt}
  M.~Cederwall, B.~E.~W.~Nilsson and D.~Tsimpis,
  %``The structure of maximally supersymmetric Yang-Mills theory:  Constraining
  %higher-order corrections,''
  JHEP {\bf 0106} (2001) 034
  [arXiv:hep-th/0102009]. 
%%%%%%%%%%%%%%%%%%%%%%%%%%%%%%%%%%%%%%%%%%%%%%%%%%%%%%%%%%%%%%%%%%%%%%%%%
%\cite{Volkov:2001tb}
\bibitem{Volkov:2001tb}
  M.~S.~Volkov,
  %``Gravitating Yang-Mills vortices in 4+1 spacetime dimensions,''
  Phys.\ Lett.\ B {\bf 524} (2002) 369
  [arXiv:hep-th/0103038].
  %%CITATION = HEP-TH 0103038;%%
%%%%%%%%%%%%%%%%%%%%%%%%%%%%%%%%%%%%%%%%%%%%%%%%%%%%%%%%%%%%%%%%%%%%%%%%%%%%%%
%\cite{Okuyama:2002mh}
\bibitem{Okuyama:2002mh}
N.~Okuyama and K.~I.~Maeda,
 %``Five-dimensional black hole and particle solution with non-Abelian gauge
%field,''
Phys.\ Rev.\ D {\bf 67} (2003) 104012
[arXiv:gr-qc/0212022].
%%CITATION = GR-QC 0212022;%%
%%%%%%%%%%%%%%%%%%%%%%%%%%%%%%%%%%%%%%%%%%%%%%%%%%%%%%%%%%%%%%%%%%%%%%%%%%%
\bibitem{Breitenlohner:1992}
P.~Breitenlohner, P.~Forgacs and D.~Maison, Nucl. Phys.
{\bf B 383} (1992) 357; {\it ibid.} {\bf 442} (1995) 126

%%%%%%%%%%%%%%%%%%%%%%%%%%%%%%%%%%%%%%%%%%%%%%%%%%%%%%%%%%%%%%%%%%%%%%%%%%%%%%
%\cite{Brihaye:2005pz}
\bibitem{Brihaye:2005pz}
  Y.~Brihaye, B.~Hartmann and E.~Radu,
  %``Deformed vortices in (4+1)-dimensional Einstein-Yang-Mills theory,''
  Phys.\ Rev.\  D {\bf 71} (2005) 085002
  [arXiv:hep-th/0502131].
  %%CITATION = PHRVA,D71,085002;%%
%%%%%%%%%%%%%%%%%%%%%%%%%%%%%%%%%%%%%%%%%%%%%%%%%%%%%%%%%%%%%%%%%%%%%%%%%%%%%%
%\cite{Hartmann:2004tx}
\bibitem{Hartmann:2004tx}
  B.~Hartmann,
  %``Non-abelian black strings,''
  Phys.\ Lett.\  B {\bf 602} (2004) 231
  [arXiv:hep-th/0409006].
  %%CITATION = PHLTA,B602,231;%%
%%%%%%%%%%%%%%%%%%%%%%%%%%%%%%%%%%%%%%%%%%%%%%%%%%%%%%%%%%%%%%%%%%%%%%%%%%%%%%
%\cite{Brihaye:2005tx}
\bibitem{Brihaye:2005tx}
  Y.~Brihaye, B.~Hartmann and E.~Radu,
  %``Black strings in (4+1)-dimensional Einstein-Yang-Mills theory,''
  Phys.\ Rev.\  D {\bf 72} (2005) 104008
  [arXiv:hep-th/0508028].
  %%CITATION = PHRVA,D72,104008;%%
%%%%%%%%%%%%%%%%%%%%%%%%%%%%%%%%%%%%%%%%%%%%%%%%%%%%%%%%%%%%%%%%%%%%%%%%%%%%%%
%\cite{Brihaye:2006ws}
\bibitem{Brihaye:2006ws}
  Y.~Brihaye and E.~Radu,
  %``Kaluza-Klein black holes with squashed horizons and d = 4 superposed
  %monopoles,''
  Phys.\ Lett.\  B {\bf 641} (2006) 212
  [arXiv:hep-th/0606228].
  %%CITATION = PHLTA,B641,212;%%
%%%%%%%%%%%%%%%%%%%%%%%%%%%%%%%%%%%%%%%%%%%%%%%%%%%%%%%%%%%%%%%%%%%%%%%%%%%%%%
%\cite{Witten:1976ck}
\bibitem{Witten:1976ck}
  E.~Witten,
  Phys.\ Rev.\ Lett.\  {\bf 38} (1977) 121.
  %%CITATION = PRLTA,38,121;%%
  %%Cited 396 times in SPIRES-HEP
%%%%%%%%%%%%%%%%%%%%%%%%%%%%%%%%%%%%%%%%%%%%%%%%%%%%%%%%%%%%%%%%%%%%%%%%%%
%\cite{Radu:2006qf}
\bibitem{Radu:2006qf}
  E.~Radu, Y.~Shnir and D.~H.~Tchrakian,
  %``Particle-like solutions to the Yang-Mills-dilaton system in d = 4+1
  %dimensions,''
  Phys.\ Rev.\  D {\bf 75} (2007) 045003
  [arXiv:hep-th/0611270].
  %%CITATION = PHRVA,D75,045003;%%
%%%%%%%%%%%%%%%%%%%%%%%%%%%%%%%%%%%%%%%%%%%%%%%%%%%%%%%%%%%%%%%%%%%%%%%%%%%%%%
%%%%%%%%%%%%%%%%%%%%%%%%%%%%%%%%%%%%%%%%%%%%%%%%%%%%%%%%%%%%%%%%%%%%%%%%%%%%%
%\cite{Maison:2004hb}
\bibitem{Maison:2004hb}
  D.~Maison,
  %``Static, spherically symmetric solutions of Yang-Mills dilaton theory,''
  Commun.\ Math.\ Phys.\  {\bf 258} (2005) 657
  [arXiv:gr-qc/0405052];
  %%CITATION = GR-QC 0405052;%%
\\
G.~Lavrelashvili and D.~Maison,
%``Static spherically symmetric solutions of a Yang-Mills field 
%coupled to a dilaton,''
Phys.\ Lett.\ B {\bf 295} (1992) 67.
%%CITATION = PHLTA,B295,67;%%
%%%%%%%%%%%%%%%%%%%%%%%%%%%%%%%%%%%%%%%%%%%%%%%%%%%%%%%%%%%%%%%%%%%%%%%%%%%%%%
%\cite{Radu:2006gg}
\bibitem{Radu:2006gg}
  E.~Radu and D.~H.~Tchrakian,
  %``Self-dual instanton and nonself-dual instanton-antiinstanton solutions in d
  %= 4 Yang-Mills theory,''
  Phys.\ Lett.\ B {\bf 636} (2006) 201
  [arXiv:hep-th/0603071].
  %%CITATION = HEP-TH 0603071;%%
%%%%%%%%%%%%%%%%%%%%%%%%%%%%%%%%%%%%%%%%%%%%%%%%%%%%%%%%%%%%%%%%%%%%%%%%%%%%%% 
 %\cite{Komar:1958wp}
\bibitem{Komar:1958wp}
  A.~Komar,
  %``Covariant conservation laws in general relativity,''
  Phys.\ Rev.\  {\bf 113} (1959) 934.
  %%CITATION = PHRVA,113,934;%%
%%%%%%%%%%%%%%%%%%%%%%%%%%%%%%%%%%%%%%%%%%%%%%%%%%%%%%%%%%%%%%%%%%%%%%%%%%%%%%
\bibitem{FIDISOL}
W. Sch\"onauer and R. Wei\ss, J. Comput. Appl. Math. {\bf 27}, 279 
(1989);
\newline
M. Schauder, R. Wei\ss\ and W. Sch\"onauer, 
{\it The CADSOL Program Package},
 Universit\"at Karlsruhe, Interner Bericht Nr. 46/92 (1992);
\newline 
W. Sch\"onauer and E. Schnepf,  ACM Trans. on Math. Soft. {\bf13}, 
333 (1987).
%%%%%%%%%%%%%%%%%%%%%%%%%%%%%%%%%%%%%%%%%%%%%%%%%%%%%%%%%%%%%%%%%%%%%%%%%%%%%%
%\cite{Belavin:1975fg}
\bibitem{Belavin:1975fg} A.~A.~Belavin, A.~M.~Polyakov, A.~S.~Shvarts and Y.~S.~Tyupkin,
 %``Pseudoparticle Solutions Of The Yang-Mills Equations,''
Phys.\ Lett.\ B {\bf 59} (1975) 85.
  %%CITATION = PHLTA,B59,85;%%
  %%Cited 672 times in SPIRES-HEP
%%%%%%%%%%%%%%%%%%%%%%%%%%%%%%%%%%%%%%%%%%%%%%%%%%%%%%%%%%%%%%%%%%%%%%%%%%%%%%


%%%%%%%%%%%%%%%%%%%%%%%%%%%%%%%%%%%%%%%%%%%%%%%%%%%%%%%%%%%%%%%%%%%%%%%%%%%%%%
\bibitem{HKK}
  B.~Hartmann, B.~Kleihaus and J.~Kunz,
  %``Axially symmetric monopoles and black holes in  Einstein-Yang-Mills-Higgs
  %theory,''
  Phys.\ Rev.\  D {\bf 65}, 024027 (2002)
  [arXiv:hep-th/0108129];
\\
    B.~Hartmann, B.~Kleihaus and J.~Kunz,
  %``Gravitationally bound monopoles,''
  Phys.\ Rev.\ Lett.\  {\bf 86}, 1422 (2001)
  [arXiv:hep-th/0009195];
\\
    B.~Kleihaus and J.~Kunz,
  %``Monopole-antimonopole solutions of Einstein-Yang-Mills-Higgs  theory,''
  Phys.\ Rev.\ Lett.\  {\bf 85}, 2430 (2000)
  [arXiv:hep-th/0006148];
\\
  Y.~Brihaye, B.~Hartmann and J.~Kunz,
  %``Gravitating dyons and dyonic black holes,''
  Phys.\ Lett.\  B {\bf 441}, 77 (1998)
  [arXiv:hep-th/9807169].
%%%%%%%%%%%%%%%%%%%%%%%%%%%%%%%%%%%%%%%%%%%%%%%%%%%%%%%%%%%%%%%%%%%%%%%%%%%%%%
\bibitem{KKS}
  B.~Kleihaus, J.~Kunz and Y.~Shnir,
  %``Gravitating monopole - antimonopole chains and vortex rings,''
   Phys.\ Rev.\ D {\bf 71} (2005) 024013
   [arXiv:gr-qc/0411106].
%%%%%%%%%%%%%%%%%%%%%%%%%%%%%%%%%%%%%%%%%%%%%%%%%%%%%%%%%%%%%%%%%%%%%%%%%%%%%%
%\cite{Kleihaus:1997mn}
\bibitem{Kleihaus:1997mn}
  B.~Kleihaus and J.~Kunz,
%   ``Static axially symmetric Einstein Yang-Mills-dilaton solutions.  I: Regular
%   solutions,''
  Phys.\ Rev.\ D {\bf 57} (1998) 834
  [arXiv:gr-qc/9707045].
  %%CITATION = GR-QC 9707045;%%
 %%%%%%%%%%%%%%%%%%%%%%%%%%%%%%%%%%%%%%%%%%%%%%%%%%%%%%%%%%%%%%%%%%%%%%%%%%%%%%
%\cite{Ibadov:2005rb}
\bibitem{Ibadov:2005rb}
  R.~Ibadov, B.~Kleihaus, J.~Kunz and M.~Wirschins,
  %``New black hole solutions with axial symmetry in Einstein-Yang-Mills
  %theory,''
  Phys.\ Lett.\  B {\bf 627} (2005) 180
  [arXiv:gr-qc/0507110].
  %%CITATION = PHLTA,B627,180;%%
%%%%%%%%%%%%%%%%%%%%%%%%%%%%%%%%%%%%%%%%%%%%%%%%%%%%%%%%%%%%%%%%%%%%%%%%%%%%%
%\cite{Ibadov:2004rt}
\bibitem{Ibadov:2004rt}
  R.~Ibadov, B.~Kleihaus, J.~Kunz and Y.~Shnir,
  %``New regular solutions with axial symmetry in Einstein-Yang-Mills  theory,''
  Phys.\ Lett.\  B {\bf 609} (2005) 150
  [arXiv:gr-qc/0410091].
  %%CITATION = PHLTA,B609,150;%%

%%%%%%%%%%%%%%%%%%%%%%%%%%%%%%%%%%%%%%%%%%%%%%%%%%%%%%%%%%%%%%%%%%%%%%%%%%%%%%
%\cite{Elvang:2003yy}
\bibitem{Elvang:2003yy}
  H.~Elvang,
  %``A charged rotating black ring,''
  Phys.\ Rev.\  D {\bf 68} (2003) 124016
  [arXiv:hep-th/0305247];
  %%CITATION = PHRVA,D68,124016;%%
\\
%\cite{Emparan:2004wy}
%\bibitem{Emparan:2004wy}
  R.~Emparan,
  %``Rotating circular strings, and infinite non-uniqueness of black rings,''
  JHEP {\bf 0403} (2004) 064
  [arXiv:hep-th/0402149].
  %%CITATION = JHEPA,0403,064;%%
%%%%%%%%%%%%%%%%%%%%%%%%%%%%%%%%%%%%%%%%%%%%%%%%%%%%%%%%%%%%%%%%%%%%%%%%%%%%%%
\end{thebibliography}
\end{document}